\documentstyle[twoside,fleqn,npb,epsfig,amsmath]{article}
\def\beq{\begin{equation}}
\def\eeq{\end{equation}}
\def\beqar{\begin{eqnarray}}
\def\eeqar{\end{eqnarray}}
\def\barr#1{\begin{array}{#1}}
\def\earr{\end{array}}
\def\bfi{\begin{figure}}
\def\efi{\end{figure}}
\def\btab{\begin{table}}
\def\etab{\end{table}}
\def\bce{\begin{center}}
\def\ece{\end{center}}

\def\text{\textstyle}

\def\al{\alpha}
\def\be{\beta}

\def\de{\delta}

\def\Si{\Sigma}

\def\De{\Delta}

\def\refeq#1{\mbox{eq.~(\ref{#1})}}
\def\refeqs#1{\mbox{eqs.~(\ref{#1})}}
\def\reffi#1{\mbox{Fig.~\ref{#1}}}

\def\refse#1{\mbox{Sect.~\ref{#1}}}

\def\citere#1{\mbox{Ref.~\cite{#1}}}
\def\citeres#1{\mbox{Refs.~\cite{#1}}}

\def\mathswitchr#1{\relax\ifmmode{\mathrm{#1}}\else$\mathrm{#1}$\fi}

\newcommand{\PW}{\mathswitchr W}
\newcommand{\PZ}{\mathswitchr Z}
\newcommand{\PA}{\mathswitchr A}

\newcommand{\PH}{\mathswitchr H}

\newcommand{\Ph}{\mathswitchr h}

\newcommand{\Pb}{\mathswitchr b}

\newcommand{\Pt}{\mathswitchr t}

\def\mathswitch#1{\relax\ifmmode#1\else$#1$\fi}

\newcommand{\MW}{\mathswitch {M_\PW}}

\newcommand{\MZ}{\mathswitch {M_\PZ}}
\newcommand{\MH}{\mathswitch {M_\PH}}

\newcommand{\Mt}{\mathswitch {m_\Pt}}

\newcommand{\mh}{\mathswitch {m_\Ph}}

\newcommand{\MA}{\mathswitch {M_\PA}}

\newcommand{\sweff}{\sin^2 \theta_{\mathrm{eff}}}

\newcommand{\GF}{\mathswitch {G_\mu}}

\def\tb{\tan\beta}

\newcommand{\Xt}{X_{\Pt}}
\newcommand{\At}{A_{\Pt}}
\newcommand{\Ab}{A_{\Pb}}

\def\order#1{${\cal O}(#1)$}
\newcommand{\mhmax}{\mh^{\rm max}}

\newcommand{\mt}{\Mt}

\newcommand{\Stop}{\tilde{t}}

\newcommand{\tsf}{\theta\kern-.20em_{\tilde{f}}}
\newcommand{\tsfp}{\theta\kern-.20em_{\tilde{f}\prime}}
\newcommand{\tsq}{\theta\kern-.15em_{\tilde{q}}}

\newcommand{\msusy}{M_{\mathrm{SUSY}}}

\newcommand{\alps}{\alpha_{\mathrm s}}

\newcommand{\feh}{{\em FeynHiggs}}

\newcommand{\cp}{{\cal CP}}

\newcommand{\VL}{\left( \begin{array}{c}}
\newcommand{\VR}{\end{array} \right)}
\newcommand{\ML}{\left( \begin{array}{cc}}
\newcommand{\MLd}{\left( \begin{array}{ccc}}
\newcommand{\MLv}{\left( \begin{array}{cccc}}
\newcommand{\MR}{\end{array} \right)}

\newcommand{\tev}{\,\, \mathrm{TeV}}
\newcommand{\gev}{\,\, \mathrm{GeV}}
\newcommand{\mev}{\,\, \mathrm{MeV}}

\newcommand{\BC}{\begin{center}}
\newcommand{\EC}{\end{center}}
\newcommand{\BE}{\begin{equation}}
\newcommand{\EE}{\end{equation}}
\newcommand{\BEA}{\begin{eqnarray}}
\newcommand{\BEAnn}{\begin{eqnarray*}}
\newcommand{\EEA}{\end{eqnarray}}
\newcommand{\EEAnn}{\end{eqnarray*}}
\newcommand{\non}{\nonumber}
\newcommand{\id}{{\rm 1\kern-.12em
\rule{0.3pt}{1.5ex}\raisebox{0.0ex}{\rule{0.1em}{0.3pt}}}}

\def\draftdate{\relax}
\def\mda{\relax}
\def\mua{\relax}
\def\mla{\relax}
\def\draft{
\def\thtystars{******************************}
\def\sixtystars{\thtystars\thtystars}
\typeout{}
\typeout{\sixtystars**}
\typeout{* Draft mode!
         For final version remove \protect\draft\space in source file
*}
\typeout{\sixtystars**}
\typeout{}
\def\draftdate{\today}
\def\mua{\marginpar[\boldmath\hfil$\uparrow$]%
                   {\boldmath$\uparrow$\hfil}%
                    \typeout{marginpar: $\uparrow$}\ignorespaces}
\def\mda{\marginpar[\boldmath\hfil$\downarrow$]%
                   {\boldmath$\downarrow$\hfil}%
                    \typeout{marginpar: $\downarrow$}\ignorespaces}
\def\mla{\marginpar[\boldmath\hfil$\rightarrow$]%
                   {\boldmath$\leftarrow $\hfil}%
                    \typeout{marginpar:
$\leftrightarrow$}\ignorespaces}
\def\Mua{\marginpar[\boldmath\hfil$\Uparrow$]%
                   {\boldmath$\Uparrow$\hfil}%
                    \typeout{marginpar: $\Uparrow$}\ignorespaces}
\def\Mda{\marginpar[\boldmath\hfil$\Downarrow$]%
                   {\boldmath$\Downarrow$\hfil}%
                    \typeout{marginpar: $\Downarrow$}\ignorespaces}
\def\Mla{\marginpar[\boldmath\hfil$\Rightarrow$]%
                   {\boldmath$\Leftarrow $\hfil}%
                    \typeout{marginpar:
$\Leftrightarrow$}\ignorespaces}
\overfullrule 5pt
\oddsidemargin -15mm
\marginparwidth 29mm
}

\allowdisplaybreaks

\title{Two-Loop Results for $\MW$ in the
Standard Model and the MSSM}

\author{
A.~Freitas,\address{Fermilab, Batavia, IL 60510-0500, USA}
S.~Heinemeyer\address{Institut f\"ur Theoretische Elementarteilchenphysik,
LMU M\"unchen, 
D--80333 Munich, Germany}
and
G.~Weiglein\address{Institute for Particle Physics Phenomenology, 
University of Durham, Durham DH1~3LE, UK}%
\thanks{Talk presented by G.~Weiglein.}
}

\begin{document}

\begin{abstract}
Recent higher-order results for the prediction of the W-boson mass,
$\MW$, within the Standard Model are reviewed and an estimate of the remaining
theoretical uncertainties of the electroweak precision observables is
given. An updated version of a simple numerical parameterisation of the
result for $\MW$ is presented.
Furthermore, leading electroweak two-loop contributions to 
the precision observables within the MSSM are discussed.
\end{abstract}

\maketitle
\vspace*{-11.9cm}
\noindent
DCPT/02/150, IPPP/02/75\\
FERMILAB-Conf-02/312-T, LMU 14/02
\vspace*{10cm}


\section{INTRODUCTION}

The comparison of electroweak precision measurements with the 
theoretical predictions allows to test the electroweak theory at the
quantum level. In this way indirect constraints on unknown parameters of
the theory can be obtained, in particular constraints on the Higgs-boson
mass, $\MH$, within the Standard Model (SM) and constraints on the
parameters of the Higgs and scalar top and bottom sector within 
the Minimal Supersymmetric extension of the SM (MSSM).

\begin{figure}[ht!]
\vspace{-2.5em}
\includegraphics[width=7.5cm,height=7.0cm]{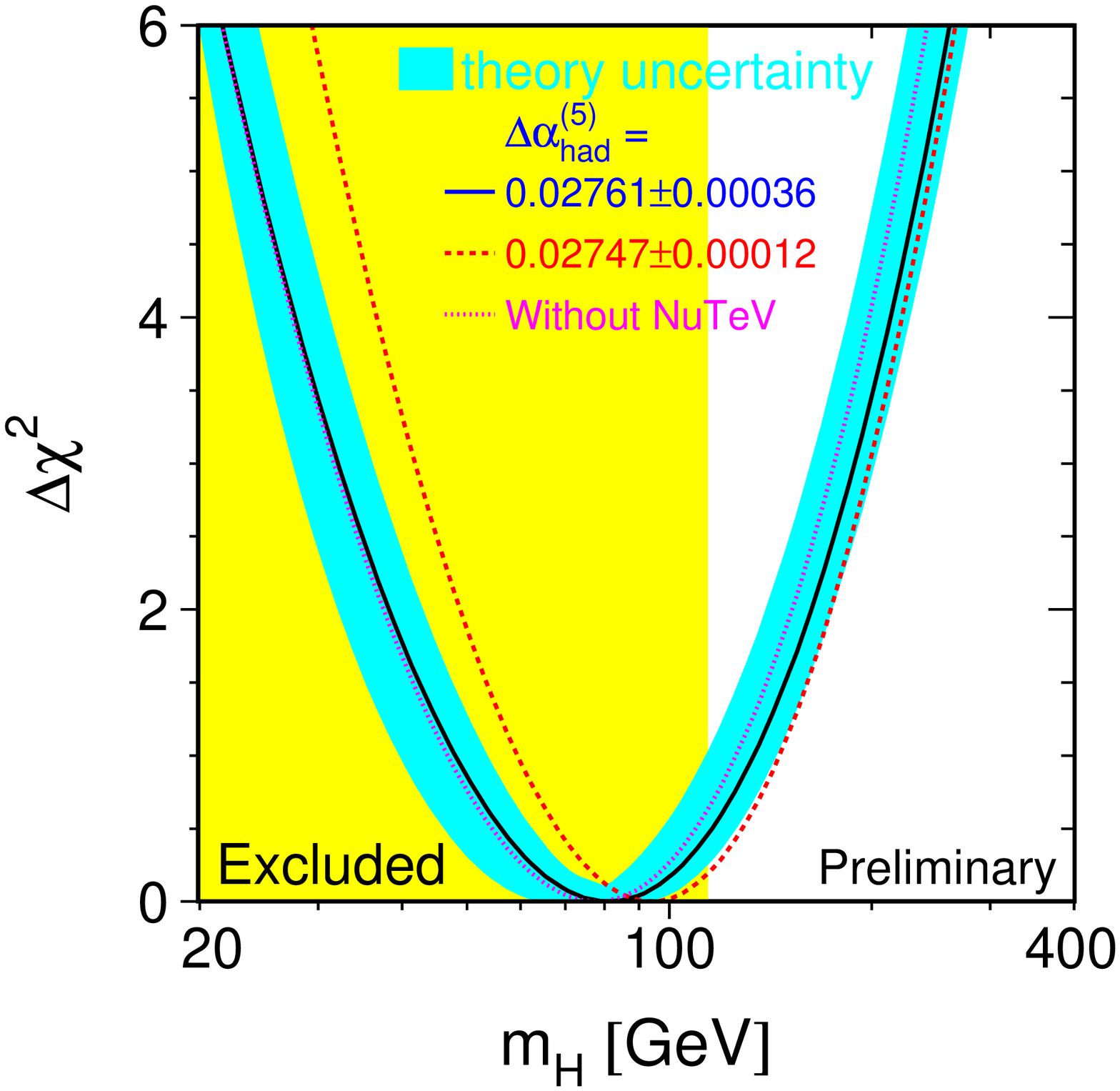}\\[-3.5em]
\includegraphics[width=7.5cm,height=7.0cm]{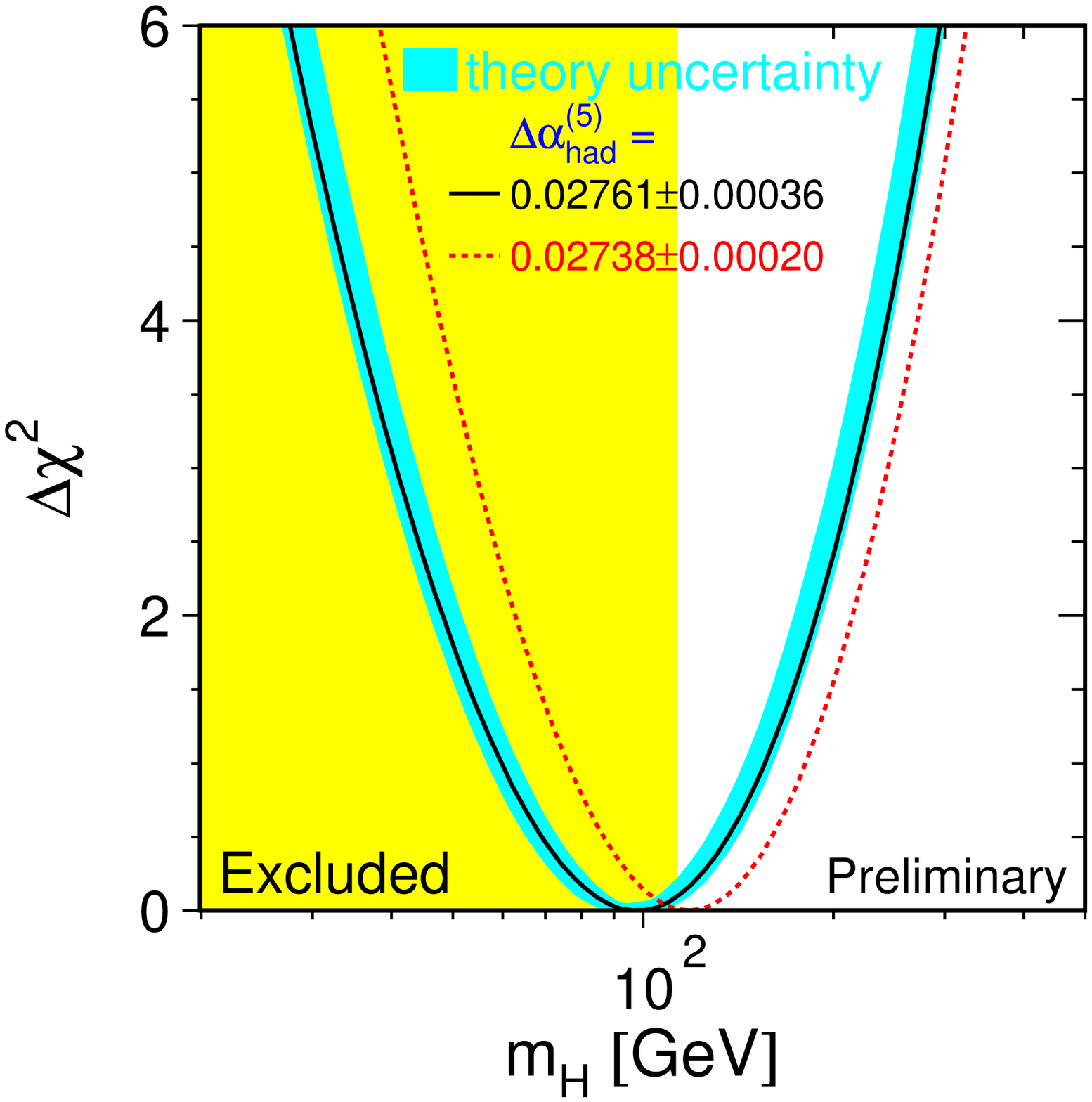}
\vspace{-5em}
\caption{Global fit to all data in the SM: comparison of the present fit
(upper plot) with the one of the winter 2001 conferences (lower plot).}
\label{fig:blueband}
\vspace{-1.7em}
\end{figure}

\reffi{fig:blueband} shows the result of a gobal fit to all data within
the SM~\cite{blueband_s02,blueband_w01}. The theoretical predictions are 
affected by two kinds of
uncertainties: the uncertainties from unknown higher-order corrections,
indicated by a ``blue band'' in \reffi{fig:blueband}, and uncertainties
from experimental errors of the input parameters, indicated in
\reffi{fig:blueband} by two fit curves corresponding to two different
values of $\De\al_{\rm had}$, the hadronic contribution to the shift in
the fine structure constant (the experimental error of the top-quark
mass, $\mt$, is directly included in the fit). The upper plot in 
\reffi{fig:blueband} shows the result based on the most recent data
(summer 2002~\cite{blueband_s02}), and the currently best estimate of the 
theoretical uncertainties from unknown higher-order corrections, while
the lower plot shows the fit result based on the
previous estimate of the theoretical uncertainties and the winter 2001
data~\cite{blueband_w01}. The comparison in \reffi{fig:blueband} shows
that the present estimate of the theoretical uncertainties from unknown
higher-order corrections yields a {\em larger\/} value than the previous
estimate. This was triggered by the recently obtained result for the 
complete fermionic two-loop corrections to the W-boson mass, $\MW$, in
the SM~\cite{MWferm1}, leading to an improved estimate of the remaining 
theroretical uncertainties in the prediction for the leptonic effective 
weak mixing angle, $\sweff$~\cite{MWferm2}.

In the following section the present status of the prediction for $\MW$
in the SM is reviewed and an estimate of the remaining theoretical
uncertainties of the electroweak precision observables is given.
\refse{sec:delromssm} summarises the impact of new electroweak two-loop 
contributions on the precision observables within the
MSSM~\cite{delrhoMSSMewnew}.

\section{PREDICTION FOR $\MW$ IN THE SM}

The prediction for $\MW$ is obtained from relating the result for the muon
lifetime within the SM (and analogously for the MSSM)
to the definition of the Fermi constant, $\GF$ 
(by convention, the QED corrections within the Fermi Model, which are
known up to two-loop order~\cite{qedfermi}, are split off in the
defining equation for $\GF$). This leads to the relation
\beq
\MW^2 \left(1 - \frac{\MW^2}{\MZ^2}\right) =
\frac{\pi \al}{\sqrt{2} \GF} \left(1 + \De r\right),
\label{eq:delr}
\eeq
where the radiative corrections are summarised
in the quantity $\De r$. The one-loop result for
$\De r$~\cite{sirlin} has first been improved by resummations of the
leading one-loop contributions from fermion loops~\cite{resum}.
Concerning irreducible two-loop contributions, the ${\cal O}(\al
\alps)$~\cite{qcd2} corrections are known for some time, while 
in the electroweak
sector results have been restricted until recently to asymptotic
expansions for large Higgs~\cite{ewmh2} and top-quark
masses~\cite{ewmt}. 

\begin{figure}[tp]
\vspace*{1mm}
\includegraphics[width=7.5cm,height=6cm]{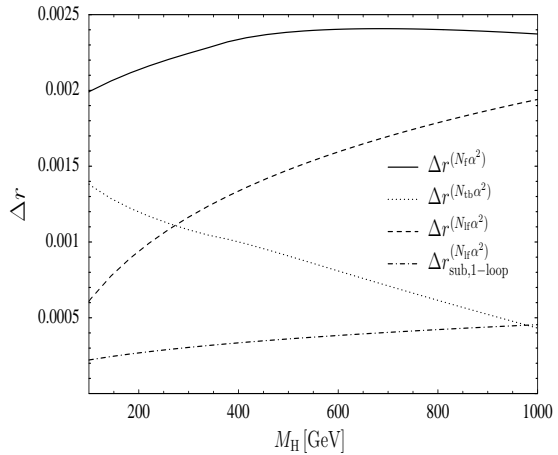}
\vspace{-3em}
\caption[]{
Relative importance of different two-loop contributions to $\De r$ with
one closed fermion loop as a function of the Higgs-boson mass, see text.
}
\label{fig:Dr2}
\vspace{-2em}
\end{figure}

Complete results for the fermion-loop contributions
at two-loop order were first obtained for the Higgs-mass dependence of
$\MW$ in \citere{ewmhdep}. The full result for the fermion-loop
contributions at two-loop order was derived in
\citeres{MWferm1,MWferm2}. \reffi{fig:Dr2} shows the relative importance
of the different contributions with one closed fermion loop to 
$\De r$ at two-loop order, whose sum is denoted by
$\De r^{(N_{\mathrm{f}} \al^2)}$. It can be seen that
both corrections with a top-/bottom-loop, 
$\De r^{(N_{\mathrm{tb}} \al^2)}$, and
with a light-fermion loop, $\De r^{(N_{\mathrm{lf}} \al^2)}$, yield
important contributions. It should be noted, however, that the
light-fermion contribution contains the numerically relatively large 
term $2 \De\al \De r^{(\al)}_{\rm bos}$, which can easily be separated
from the genuine two-loop contribution of the light fermions. In order
to investigate the numerical relevance of the latter, in \reffi{fig:Dr2}
also the difference $\Delta r^{(N_{\rm lf} \alpha^2)}_{\rm sub,1-loop} =
\De r^{(N_{\mathrm{lf}} \al^2)} - 2 \De\al \De r^{(\al)}_{\rm bos}$ is shown.
While these genuine light-fermion two-loop contributions do not exceed
the top-/bottom contributions for any value of the Higgs-boson
mass below 1~TeV, 
they nevertheless amount up to $3.3 \times 10^{-4}$, which corresponds to
a shift in $\MW$ of more than 5~MeV.

Recently also the Higgs-mass dependence
of the purely bosonic two-loop corrections became
available~\cite{MWferm2}. Finally, the full result for the purely
bosonic two-loop corrections has been obtained in \citere{MWbos},
completing in this way the calculation of muon decay at the two-loop
level. The numerical effect of the purely bosonic two-loop corrections
turned out to be relatively small, giving rise to a shift in $\MW$ of
less than $\pm 1$~MeV for $\MH \leq 1$~TeV. From the higher-order contributions
to $\De r$ (for a discussion, see e.g.\ \citere{MWferm2}) only the
top-bottom contributions at ${\cal O}(\al \alps^2)$~\cite{qcd3} were
found to be non-negligible in view of the present experimental
accuracies.

Below a simple parameterisation of the result for $\MW$ is given, being 
based on taking into account the following contributions to $\De r$, 
\beqar
\De r &=& \De r^{(\al)} + \De r^{(\al\alps)} + \De r^{(\al\alps^2)} \non \\
&& \mbox{} + \De r^{(N_{\mathrm{f}} \al^2)} 
   + \De r^{(N_{\mathrm{f}}^2 \al^2)} + \De r^{(\al^2, \mathrm{bos})},
\non \\
&& 
\label{eq:delrcontribs}
\eeqar
where $\De r^{(\al)}$ is the one-loop result, $\De r^{(\al\alps)}$ and 
$\De r^{(\al\alps^2)}$ are the two-loop and three-loop QCD corrections, 
$\De r^{(N_{\mathrm{f}} \al^2)}$ and $\De r^{(N_{\mathrm{f}}^2 \al^2)}$
are the electroweak two-loop contributions with one and two fermion
loops, respectively, and $\De r^{(\al^2, \mathrm{bos})}$ is the purely
bosonic two-loop contribution according to the expression given in 
\citere{MWbos2}. The numerically rather small 
electroweak higher-order corrections have been neglected
here. The parameterisation of the result for $\MW$ reads
\beqar
\MW &=& \MW^0 - d_1 \, \mathrm{dH} - d_2 \, \mathrm{dH}^2 
              + d_3 \, \mathrm{dH}^4 \non \\
&& \mbox{} - d_4 \, \mathrm{d}\al + d_5 \, \mathrm{dt} 
           - d_6 \, \mathrm{dt}^2 - d_7 \, \mathrm{dH} \, \mathrm{dt} \non \\
&& \mbox{} - d_8 \, \mathrm{d}\alps + d_9 \, \mathrm{dZ} ,
\label{eq:cplxpar}
\eeqar
where the dependence on the variables 
$\MH$, $\mt$, $\al$, $\alps$ and $\MZ$ is expressed by
${\rm dH} = \ln\left(\MH/(100 \gev)\right)$, 
${\rm dt} = \left(\mt/(174.3 \gev)\right)^2 - 1$,
${\rm d}\al = \De\al/0.05924 - 1$,
${\rm d}\alps = \alps(\MZ)/0.119 - 1$, and
${\rm dZ} = \MZ/(91.1875 \gev) -1$.
The coefficients $d_1, \ldots, d_9$ have the following values (in GeV)
\beq
\begin{array}{rclrcl}
\MW^0 &=& 80.3757 , &
d_5 &=& 0.5236 , \\
d_1 &=& 0.05515 , &
d_6 &=& 0.0727 , \\
d_2 &=& 0.009803 , &
d_7 &=& 0.00541 , \\
d_3 &=& 0.0006078 , \; &
d_8 &=& 0.0765 , \\
d_4 &=& 1.078 , &
d_9 &=& 115.0 .
\end{array}
\eeq
Employing these coefficients, the simple parameterisation of 
\refeq{eq:cplxpar} approximates the full result for $\MW$ based on
the contributions given in \refeq{eq:delrcontribs} with an accuracy of 
better than $0.3$~MeV for $65 \gev \leq$~$\MH$~$\leq 1 \tev$ and 
$2\sigma$ variations of all other experimental input values. This
formula, which includes the recently obtained result for 
$\De r^{(\al^2, \mathrm{bos})}$~\cite{MWbos,MWbos2}, updates the
parameterisation given in \citere{MWferm1}. As discussed above, the
corresponding shift in $\MW$ lies within about $\pm 1$~MeV.

The remaining theoretical uncertainties of the electroweak precision
observables from unknown higher-order
corrections, taking into account all known contributions, can be
estimated with the methods described in \citeres{MWferm2,snowmass} as:
\beq
\de \MW^{\rm th} \approx \pm 6 \mev, \quad
\de \sweff^{\rm th} \approx \pm 7 \times 10^{-5}.
\label{eq:unchighord}
\eeq
They are smaller at present than the parametric uncertainties 
from the experimental errors of the input parameters $\mt$ and
$\De\al_{\rm had}$. The experimental errors of $\de \mt = \pm
5.1$~GeV~\cite{blueband_s02} and 
$\de(\De\al_{\rm had}) = 36 \times 10^{-5}$~\cite{blueband_s02}
induce theoretical uncertainties of 
\beqar
\de \MW^{\rm th} \approx \pm 31 \mev, && \hspace{-1.5em}
     \de \sweff^{\rm th} \approx \pm 16 \times 10^{-5}, \hspace{1.5em} \non \\
\de \MW^{\rm th} \approx \pm 6.5 \mev, && \hspace{-1.5em}
     \de \sweff^{\rm th} \approx \pm 13 \times 10^{-5}, \hspace{1.5em}
\eeqar
respectively. For comparison, the present experimental errors of $\MW$
and $\sweff$ are~\cite{blueband_s02}
$$
\de \MW^{\rm exp} \approx \pm 34 \mev, \quad
\de \sweff^{\rm exp} \approx \pm 17 \times 10^{-5}.
$$
At the next generation of colliders, i.e.\ RunII of the Tevatron, the
LHC and an $e^+e^-$ Linear Collider running at the Z-boson resonance and
the WW-threshold, these experimental errors will be reduced to
about (see \citere{snowmass} and references therein)
\beq
\de \MW^{\rm exp} \approx \pm 6-7 \mev, \quad
\de \sweff^{\rm exp} \approx \pm 1 \times 10^{-5}.
\label{eq:experrfut}
\eeq
At the same time, improved measurements will also
reduce the parametric uncertainty from the experimental errors of the
input parameters to about~\cite{snowmass,gigaz}.
\beq
\de \MW^{\rm th} \approx \pm 2 \mev, \quad
\de \sweff^{\rm th} \approx \pm 2 \times 10^{-5}.
\label{eq:partheoerrfut}
\eeq
Further work on higher-order corrections will clearly be needed in order to
reduce the uncertainties from unknown higher-order corrections below the
level of \refeqs{eq:experrfut}, (\ref{eq:partheoerrfut}).

\section{LEADING ELECTROWEAK 2-LOOP CORRECTIONS IN THE MSSM}
\label{sec:delromssm}

The situation concerning theoretical uncertainties of the electroweak
precision observables $\MW$ and $\sweff$ from unknown higher-order
corrections within the MSSM is significantly worse than in the SM.
Comparing the available results for higher-order corrections in both
models, the uncertainties from unknown higher-order corrections within
the MSSM can be estimated to be at least a factor of 2 larger than the
ones in the SM as given in \refeq{eq:unchighord}.

The leading higher-order corrections from quark and squark loops enter
via the quantity $\De\rho$,
\beq
\De\rho = \frac{\Si_{\PZ}(0)}{\MZ^2} - \frac{\Si_{\PW}(0)}{\MW^2} ,
\label{delrho}
\eeq
where $\Si_{\PZ,\PW}(0)$ denote the transverse parts of the unrenormalised
Z- and W-boson self-energies at zero momentum transfer, respectively.
Within the MSSM, the two-loop corrections of \order{\al\alps} to 
$\De\rho$~\cite{dr2lA} as well as the gluonic two-loop corrections to
$\De r$~\cite{dr2lB} have been obtained. Concerning electroweak two-loop
corrections, in the limit of a large SUSY scale,
$\msusy \gg \MZ$, where the SUSY particles decouple, the contributions in 
the MSSM reduce to those of a Two-Higgs-Doublet model with MSSM
restrictions. As a first result in this context, the \order{\al_t^2}
corrections in the limit where the lightest $\cp$-even
Higgs boson mass vanishes, i.e.\ $\mh \to 0$, have been obtained in
\citere{drMSSMgf2}. Recently the \order{\al_t^2}, \order{\al_t \al_b}
and \order{\al_b^2} contributions to $\De\rho$ for $\msusy \gg \MZ$ have
been evaluated for arbitrary values of $\mh$~\cite{delrhoMSSMewnew}. As in
the case of the SM, the numerical effect of going to non-vanishing values 
of the Higgs-boson mass turned out to be sizable.

\begin{figure}[ht!]
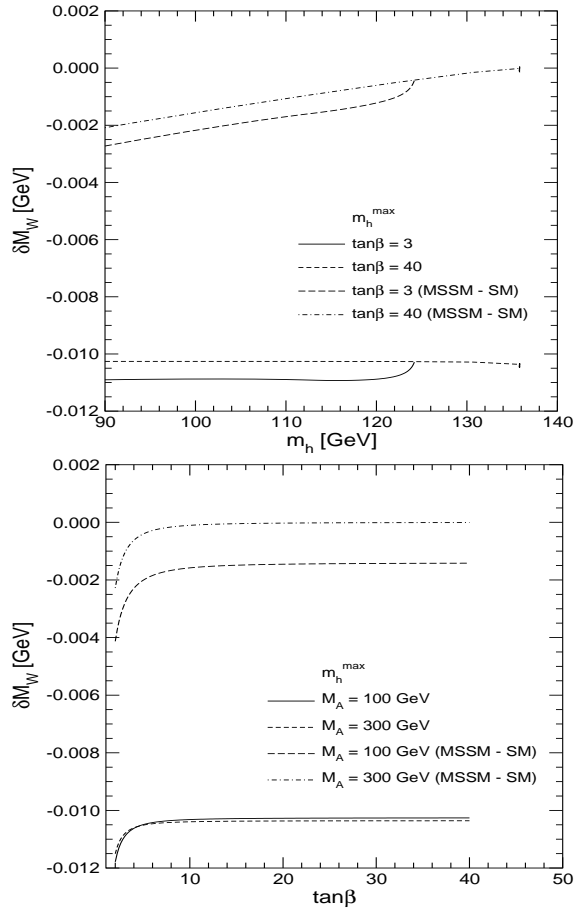

\vspace{1.5mm}
\includegraphics[width=7.5cm,height=6.0cm]{delrhoMT2Yukfull44.bw.eps}\\
\includegraphics[width=7.5cm,height=6.0cm]{delrhoMT2Yukfull54.bw.eps}
\vspace{-3em}
\caption{Contribution of the \order{\al_t^2} MSSM corrections to $\MW$
as a function of $\mh$ (upper plot) and $\tan\be$ (lower plot).}
\label{fig:delMWMSSM}
\vspace{-2em}
\end{figure}

In \reffi{fig:delMWMSSM} the numerical effect of the \order{\al_t^2}
corrections on $\MW$ is analysed. In addition to the MSSM
\order{\al_t^2} correction to $\de\MW$ also 
the `effective' change from the SM result (where the value of the 
SM Higgs boson mass
has been set to $\mh$) to the new MSSM result is shown.
The parameters in \reffi{fig:delMWMSSM} are chosen according to the 
$\mhmax$~benchmark scenario~\cite{LHbenchmark}, i.e.\ 
$\msusy = 1$~TeV, $\Xt = 2\, \msusy$, where $\mt\Xt$
is the off-diagonal entry in the $\Stop$~mass matrix.
The other parameters are $\mu = 200 \gev$, $\Ab = \At$. The Higgs-boson
mass $\mh$ is
obtained in the upper plot from varying $\MA$ from
$50 \gev$ to $1000 \gev$, while keeping $\tb$ fixed at $\tb = 3, 40$.
In the lower plot, $\tb$ is varied from 2 to 40, $\MA$ is kept fixed at
$\MA = 100, 300 \gev$. The calculation of $\mh$ from the other MSSM
parameters contains corrections up to two-loop order, as implemented in
the program \feh~\cite{feynhiggs}.

The effect of the \order{\al_t^2} MSSM contributions on $\de\MW$
amounts up to $-12 \mev$. For large 
$\tb$ it saturates at about $-10 \mev$.
The `effective' change in $\MW$ in comparison with the corresponding 
SM result with
the same value of the Higgs-boson mass is significantly smaller. It amounts 
up to $-3 \mev$ and goes to zero for large $\MA$ as expected from the
decoupling behaviour.
For a small $\cp$-odd Higgs boson mass,
$\MA = 100 \gev$, a shift of $-2 \mev$ in $\MW$ remains also in the
limit of large $\tb$, since the two Higgs doublet sector does not
decouple from the MSSM. For large $\MA$, $\MA = 300 \gev$, for nearly
all $\tb$ values the effective change in $\MW$ is small.

The absolute contribution for $\de\sweff$ (which is not shown here) 
is around $+6 \times 10^{-5}$. The effective change ranges between
$+3 \times 10^{-5}$ for small $\tb$ and small $\MA$ and
approximately zero for large $\tb$ and large $\MA$.


\smallskip
\noindent
{\em Acknowledgements:} G.W.\ thanks the organisers of ``RADCOR 2002 -- 
Loops \& Legs 2002'' for the invitation and the pleasant atmosphere at
the meeting.
This work has been supported by the European Community's Human
Potential Programme under contract HPRN-CT-2000-00149 Physics at
Colliders.


\end{document}